\documentclass[twocolumn,showpacs,preprintnumbers,amsmath,amssymb]{revtex4-1}

\usepackage{graphicx}
\usepackage{dcolumn}
\usepackage{bm,color}

\newcommand{\Eref}[1]{eq.(\ref{#1})}
\newcommand{\Fref}[1]{Fig.\ref{#1}}
\newcommand{\figwidth}{3.35in}
\newcommand{\nin}{\in \hspace{-.37em}/ }
\newcommand{\ninn}{\in \hspace{-.90em}/ }

\newcommand{\simgeq}{\hspace{0.3em}\raisebox{0.4ex}{$>$}\hspace{-0.75em}\raisebox{-.7ex}{$\sim$}\hspace{0.3em}}

\begin{document}

\preprint{APS/123-QED}

\title{A Mean-field Approach for an Intercarrier Interference Canceller for OFDM}

\author{Ayaka Sakata }
\email{ayaka@sp.dis.titech.ac.jp}
\author{Yoshiyuki Kabashima}
\affiliation{Department of Computational 
Intelligence and Systems Science, Tokyo Institute of Technology, 
Midori-ku, Yokohama 226-8502, Japan.}
\author{Yitzhak Peleg}
\affiliation{Department of Physics,
Bar-Ilan University, \\
Ramat-Gan 52900, Israel.}

\date{\today}

\begin{abstract}
The similarity of the mathematical description of
random-field spin systems to orthogonal 
frequency-division multiplexing (OFDM) scheme for
wireless communication is exploited in 
an intercarrier-interference 
(ICI) canceller used
in the demodulation of OFDM. 
The translational symmetry in the Fourier domain 
generically concentrates the major contribution of ICI from each subcarrier in 
the subcarrier's neighborhood. 
This observation in conjunction with mean field approach leads to a
development of an
ICI canceller whose necessary cost of 
computation scales 
linearly with respect to the number of subcarriers. 
It is also shown that
the dynamics of the mean-field canceller are well captured by a discrete map of 
a single macroscopic variable,
without taking the spatial and time correlations of estimated variables into account.
\end{abstract}


\maketitle

\section{Introduction}

Wireless communication technologies play a significant role
in the modern information society. 
As of the end of 2010, 
there are more than 4.6 billion mobile-cellular 
subscriptions in the world \cite{UNNews},
and the use of wireless devices (such as personal 
digital assistants and GPS units) is ever increasing. 
To keep up with the accompanying rapid growth in data traffic,
communication efficiency of today's wireless communication systems
must thus be constantly improved.

A decade has passed since a fruitful connection between 
wireless communications and statistical mechanics
was introduced by a seminal work by Tanaka \cite{Tanaka}.
On the basis of an analogy
between the demodulation problem of wireless 
communications and statistical mechanics
of disordered Ising spin systems,  
he successfully clarified the potential efficiency 
of a wireless communication scheme
known as code-division multiple access
(CDMA), which is employed in the third generation 
cellular phone systems.  
Later, this analogy was also utilized in developing 
practically feasible and efficient demodulation algorithms
for CDMA \cite{Kabashima,Tanaka-Okada}.

In a recent study \cite{Efraim}, the connection to statistical mechanics 
was extended to another wireless-communication 
scheme, namely, orthogonal frequency-division multiplexing
(OFDM), which is today employed in the fourth generation cellular phones 
and the latest Wi-Fi systems.  
According to this scheme, the available frequency domain is divided
into sub-domains, and data is transmitted by the subcarriers
associated with those sub-domains \cite{Chang}.
Because of the orthogonality 
between the subcarriers, they
can be closely placed
in the frequency domain, 
thereby attaining 
high-rate-data transmission.
However, in a mobile radio environment, relative
movement brings about a Doppler spread, which destroys
the orthogonality between the subcarriers. 
This destruction leads to the occurrence of intercarrier interference (ICI),  
which rapidly deteriorates the bit error rate. 
An efficient ICI-cancellation scheme is, therefore, indispensable 
in use of the OFDM scheme in such environments. 
In \cite{Efraim}, a Monte-Carlo-based ICI cancellation scheme
was developed on the basis of mapping 
an OFDM model
to a variant of random-field Ising spin systems. 
Numerical experiments indicated that 
the developed scheme can achieve significantly 
better performance than existing standard 
methods in terms of bit error rate. 
However, the cost of computation, which grows with the 
square of the number of subcarriers, and the technical 
difficulty in implementing electrical circuits prevent 
the scheme from being practically significant. 

The purpose of this study is to develop an approximate ICI-cancellation 
scheme 
for resolving the above-mentioned drawbacks. 
For this purpose, 
a mean-field approximation (MFA) is utilized 
in conjunction with
the analogy between OFDM and random field Ising spin systems.
Naive MFA requires as much computational 
cost as 
the Monte-Carlo based cancellation scheme. 
We show that utility of translational 
invariance in the Fourier domain, which is intrinsic 
in OFDM, makes it possible to develop
an ICI cancellation scheme whose computational 
cost is proportional to the number of subcarriers.
We also show that the performance of the developed 
algorithm based on MFA
is well captured by a discrete map of a single 
variable and that the fact is supported by numerical experiments.

This paper is organized as follows.
In section \ref{model}, we explain the OFDM model
studied in this paper.
In section \ref{modulation}, we propose a
MFA-based ICI cancellation scheme.
In section \ref{results}, we explore the performance of the proposed method
and its time evolution, and derive the approximated expression of the 
proposed canceller. The derivation of the approximated expression is 
explained in Appendix \ref{variance_derivation}.
Finally, section \ref{summary} is devoted to the conclusion and summary.

\section{Model}
\label{model}

When a time sequence of a signal, $\bm{x}=\{x_t\}~(t=1,\cdots,N)$, is transmitted in a mobile radio environment, 
the received symbol, $\bm{y}=\{y_t\}$, in the multipath channel is expressed by
\begin{align}
y_t=\frac{1}{\sqrt{M}}\sum_{p=1}^Mh_p\exp\Big(\sqrt{-1}\frac{2\pi\epsilon_p t}{N}\Big)x_t+\eta_t.
\label{time_response}
\end{align}
The time delay of each path is assumed to be zero for simplicity, and the channel noise, $\{\eta_t\}$, 
is independent with respect to the time domain. 
The number of paths is $M$, 
the amplitude of each path, $\{h_p\}$, 
is distributed according to
the Rayleigh distribution \cite{Zhao1,Zhao2} as
\begin{align}
P(h_p)=h_p\exp(-\frac{h_p^2}{2}),
\end{align}
and $\{\epsilon_p\}$ 
is the Doppler shift, which is assumed to be distributed 
uniformally in the region $\epsilon_p\in[0,\epsilon_{\max}]$.
With this model, the difference between the maximum and minimum values 
of the Doppler shift is significant 
irrespective of the sign of the Doppler shift.

By applying the discrete Fourier transformation to \Eref{time_response}, 
the frequency-domain representation of the transmitted signal, $\bm{X}=\{X_k\}$, and the received signal, $\bm{Y}=\{Y_k\}$, 
where $k=0,\cdots,N-1$ and $N$ is the number of subcarriers, is given by
\begin{align}
Y_k=\sum_{l=1}^{N}W_{kl}X_l+n_k, 
\end{align}
where $n_k$ is the discrete Fourier transform of the channel noise.
We assume that $n_k$ is characterized as an additive white Gaussian noise (AWGN) 
of mean zero and variance $\sigma_0^2$. 
The component of $N\times N$ matrix $\bm{W}$, 
called a frequency-domain matrix, is given by
\begin{align}
W_{kl}=\sum_{p=1}^M\frac{h_p\sin(\pi(l-k+\epsilon_p))e^{\sqrt{-1}(1-\frac{1}{N})\pi(l-k+\epsilon_p)}}{\sqrt{M}N\sin(\pi(l-k+\epsilon_p)\slash N)},
\end{align}
where $W_{kl}$ represents the intensity of the interference from subcarrier $l$ to $k$. 
When $\{\epsilon_p\}=\bm{0}$, the matrix is diagnonal, namely 
$W_{kl}=\delta_{k,l}$, 
where $\delta$ is Kronecker's delta,
and there is no ICI between any subcarriers. 
In general, the frequency-domain matrix has 
translation symmetry, 
so the value of each component 
$W_{kl}$ only depends on the difference of the indices, namely $k-l$. 
This fact is a result of the Fourier representation.

The Doppler shift, $\epsilon_p$, is normalized by the
frequency separation of subcarriers, $\Delta f=1\slash N$, as 
$\epsilon_p=f_{D,p}\slash \Delta f$, where $f_{D,p}$ is the
Doppler frequency at $p$-th path.
The parameter $\epsilon_p$ indicates the influence of the Doppler effect
on ICI with a given alignment of subcarriers.

The transmitted bits, channel noise, and received bits are represented as complex numbers. 
Their real and imaginary parts are denoted by $\bm{X}^{\rm R}$ and $\bm{X}^{\rm I}$, 
$\bm{n}^{\rm R}$ and $\bm{n}^{\rm I}$, and $\bm{Y}^{\rm R}$ and $\bm{Y}^{\rm I}$, 
respectively. Hereafter, they are represented as vectors consisting of $2N$ elements: 
$\bm{X}\equiv[\bm{X}^R,\bm{X}^I]^T$, $\bm{n}\equiv[\bm{n}^{\rm R},\bm{n}^{\rm I}]^T$, 
and $\bm{Y}\equiv[\bm{Y}^R,\bm{Y}^I]^T$, respectively. 
$T$ denotes the operation of the matrix transpose. 
The corresponding frequency-domain 
matrix is redefined as a $2N\times  2N$ matrix, 
$\bm{W}\equiv 
\begin{bmatrix}
\bm{W}^R  & -\bm{W}^I\\
\bm{W}^I  & \bm{W}^R
\end{bmatrix}
$, 
where $\bm{W}^R$ and $\bm{W}^I$ are the real and imaginary parts of the matrix, respectively 
\cite{Efraim}.

\section{Mean-field ICI canceller}
\label{modulation}

The problem of the OFDM system is to recover the original signal
by canceling out the inter-carrier interference.
The Bayesian framework offers various ICI canceling strategies on the basis of the posterior distribution, 
\begin{align}
P(\bm{X}|\bm{Y})=\frac{P(\bm{Y}|\bm{X})P(\bm{X})}{\sum_{\bm{X}}P(\bm{Y}|\bm{X})P(\bm{X})},
\end{align}
where the likelihood $P(\bm{Y}|\bm{X})$ is given by the distribution of 
the channel noise \cite{Li}. 
At the receiving side, 
it is assumed that the channel noise
is described by the Gauss distribution with mean zero and variance $\sigma^2$, 
\begin{align}
P(\bm{Y}|\bm{X})=\Big(\frac{1}{2\pi\sigma^2}\Big)^{N\slash 2}\exp\Big(-\frac{(\bm{W}\bm{X}-\bm{Y})^2}{2\sigma^2}\Big),
\end{align}
and prior distribution $P(\bm{X})$ is the uniform distribution. 
The posterior probability can therefore be expressed as
\begin{align}
\nonumber
P(\bm{X}|\bm{Y})&=\frac{1}{Z}\exp\Big\{\frac{1}{\sigma^2}\Big(\frac{1}{2}\sum_{i,j}J_{ij}X_iX_
 j-\sum_{i=1}^{2N}h_iX_i\Big)\Big\}\\
&\equiv\frac{1}{Z}e^{-\beta{\cal H}(\bm{X}|\bm{J},\bm{h})}
\label{H_def}
\end{align}
where
$\sigma^{-2}$ is identified with 
the {``inverse temperature''} $\beta$,
the interaction matrix and the external field are given by 
$\bm{J}=\bm{W}^{T}\bm{W}$ and $\bm{h}=\bm{Y}^{T}\bm{W}$, respectively,
and $Z$ corresponds to the partition function.
The interaction matrix,
$\bm{J}= 
\begin{bmatrix}
\bm{J}_1  & \bm{J}_2\\
\bm{J}_2^T  & \bm{J}_1
\end{bmatrix}
$, 
also has translational invariance; 
the $(k,l)$-components of $N\times N$ sub-matrices $\bm{J}_1$ and 
$\bm{J}_2$ only depend on $k-l$.
%

The Hamiltonian of the OFDM system, ${\cal H}(\bm{X}|\bm{J},\bm{h})$,
defined in \Eref{H_def},
can be regarded as that for a random-field spin model.
Unlike typical random field models,
the random field of the OFDM model 
is determined by the ICI between transmitted bits
and the channel's properties.
In particular, when the number of the paths, $M$, is equal to 1,
the value of the off-diagonal element
is much smaller than that of the diagonal element;
hence, the system can be regarded as a single-body problem
with random fields.

The maximum a posterior probability (MAP) strategy 
\begin{align}
\hat{\bm{X}}={\rm arg}\max_{\bm{X}}P(\bm{X}|\bm{Y}),
\label{eq:MAP}
\end{align}
where ${\rm arg}\max$ is the argument giving the maximum value of the function, 
is guaranteed to minimize the block-wise error probability. 
This strategy corresponds to the search of the ground state of Hamiltonian 
${\cal H}(\bm{X}|\bm{J},\bm{h})$.
However, the numerical cost of exactly obtaining the maximizer $\hat{\bm{X}}$ 
increases exponentially with increasing
number of subcarriers. 
Zero temperature ($\beta \to \infty$)
synchronous dynamics of $N$ symbols 
based 
on the
mean-field approximation (MFA)
is a practically feasible approximate scheme for finding the MAP solution of \Eref{eq:MAP}
\cite{Varanasi}. 
In the case of the quadrature-phase-shift-keying (QPSK) modulation, 
where the components of $\bm{X}$ takes one of the two values, $\pm 1$,
the scheme is expressed as
\begin{align}
\hat{X}_k^{t+1}={\rm sgn}(h_k-\sum_{l\neq k}J_{kl}\hat{X}_l^t),
\label{decode_full}
\end{align}
where 
${\rm sgn}(u)$ denotes the sign of $u$ and
$\hat{X}_k^t$ is the tentative decision of the $k$-th symbol after $t$ 
iterations. 
It is assumed that the configuration $\{\hat{X}_k^t\}$ is invariant at 
$\{\hat{X}_k^*\}$ after sufficient updates, and the fixed configuration 
is regarded as the final decision of the transmitted bits. 
Synchronous update schemes similar to \Eref{decode_full} have been introduced 
for evaluating the minimum mean square error (MMSE) estimator of 
Gaussian priors $P({\bm X}) \propto \exp\left (-|{\bm X}|^2/(2 \sigma_X^2) \right )$,  
in which the transmitted bits are estimated as
$\hat{\bm X}_{\rm MMSE}=\left ((\sigma^2/\sigma_X^{2}) {\bm I} +\bm J\right )^{-1}{\bm h}$
in the current case, 
where ${\bm I}$ is the identity matrix \cite{Goro,Hou,Molisch}.

The computational cost of \Eref{decode_full}
increases as $O(N^2)$, 
and it may reduce the practical feasibility of \Eref{decode_full}. 
To reduce this cost increase,
the proposed algorithm utilizes the fact that 
the absolute value of ${J_1}_{kl}$ and ${J_2}_{kl}$ decreases as $|k-l|$ increases, 
which indicates that the major contributions to ICI
(to which each subcarrier is subject) are concentrated on 
the subcarrier's neighborhood. 
Based on this observation, 
the strategy proposed here considers only a part of 
the ICI among subcarriers in the frequency domain at each stage of the cancellation. 
A similar strategy was also proposed for a Gaussian MMSE estimator \cite{Schniter}.

Let us define the set of indices of subcarriers that are 
considered to be contributed on ICI of subcarrier 
$k$ as 
$\partial_k(\omega)
\equiv\{x,x+N|x={\rm mod}(k+N-1\pm 
y,N)+1,y\in\mathbb{N}, 1\leq y\leq\omega\}$,
where ${\rm mod}(k,N)$ means the remainder of $k\slash N$ as an integer. 
In general, $\partial_k(\omega)$ is composed of $4 \omega$ elements; 
for instance,  
$\partial_1(1)=\{2, N, N+2, 2N\}$.
The update rule is then given as
\begin{align}
\hat{X}_k^{t+1}={\rm sgn}(h_k-\sum_{l\in\partial_k(\omega)}J_{kl}\hat{X}_l^t).
\label{decode_w}
\end{align}
The algorithm 
(\ref{decode_full}) 
corresponds to the case that $\omega=\lfloor N\slash 2\rfloor$ in \Eref{decode_w},
where $\lfloor x\rfloor$ denotes the largest integer not greater than
real number $x$.
The numerical cost per iteration of \Eref{decode_w} is $O(N)$ as long as parameter $\omega$ is $O(1)$. 
When $\omega<N\slash 2$, fixed point $\{\hat{X}_i^*\}$ does not correspond to 
the maximizer of the posterior probability,
but it is expected to be a good approximation of the maximizer.

\section{Results}
\label{results}

\subsection{Performance of ICI canceller}

We observe bit error rate (BER), which is defined by
\begin{align}
{\rm BER}=\frac{1}{2N}\sum_{i=1}^{2N}\langle\overline{\hat{X}_i^*X_i}\rangle,
\end{align}
where $\overline{\cdots}$ and $\langle\cdots\rangle$ represent the average over the 
frequency-domain matrix 
$\bm{W}$ and over the channel noise and transmitted symbol, respectively.
The BER performance of the decoder is bounded from below by that
for 
a single bit transmitted through 
the AWGN channel,
which is given by
\begin{align}
{\rm BER}^{\rm OPT}=\frac{1}{2}{\rm erfc}(\sqrt{{\rm Eb}/{\rm N0}}),
\end{align}
because there is no ICI involved 
when only a single bit is transmitted.
BER of the ICI canceller closes to 
${\rm BER}^{\rm OPT}$
as the elimination of ICI becomes successful.
In 
the current
model, the signal to noise ratio (SNR) is given by,
\begin{align}
{\rm SNR}=\frac{1}{N}\sum_{i,j=1}^{N}\frac{|W_{ij}|^2}{\sigma_0^2}.
\end{align}
In the QPSK modulation case, 
the number of bits per symbol is two, so
Eb$/$N0 corresponds to SNR$/$2 \cite{Zhao1}.

\begin{figure}
\begin{center}
\includegraphics[width=\figwidth]{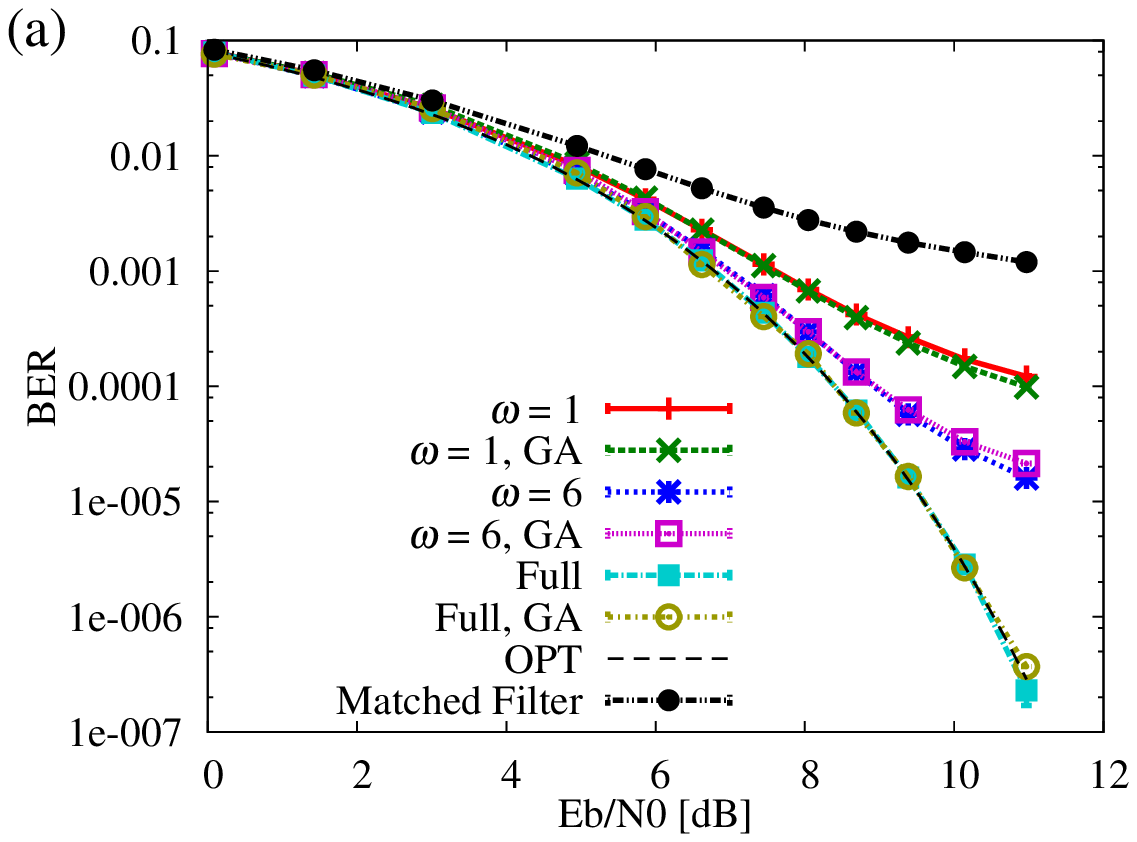}
\includegraphics[width=\figwidth]{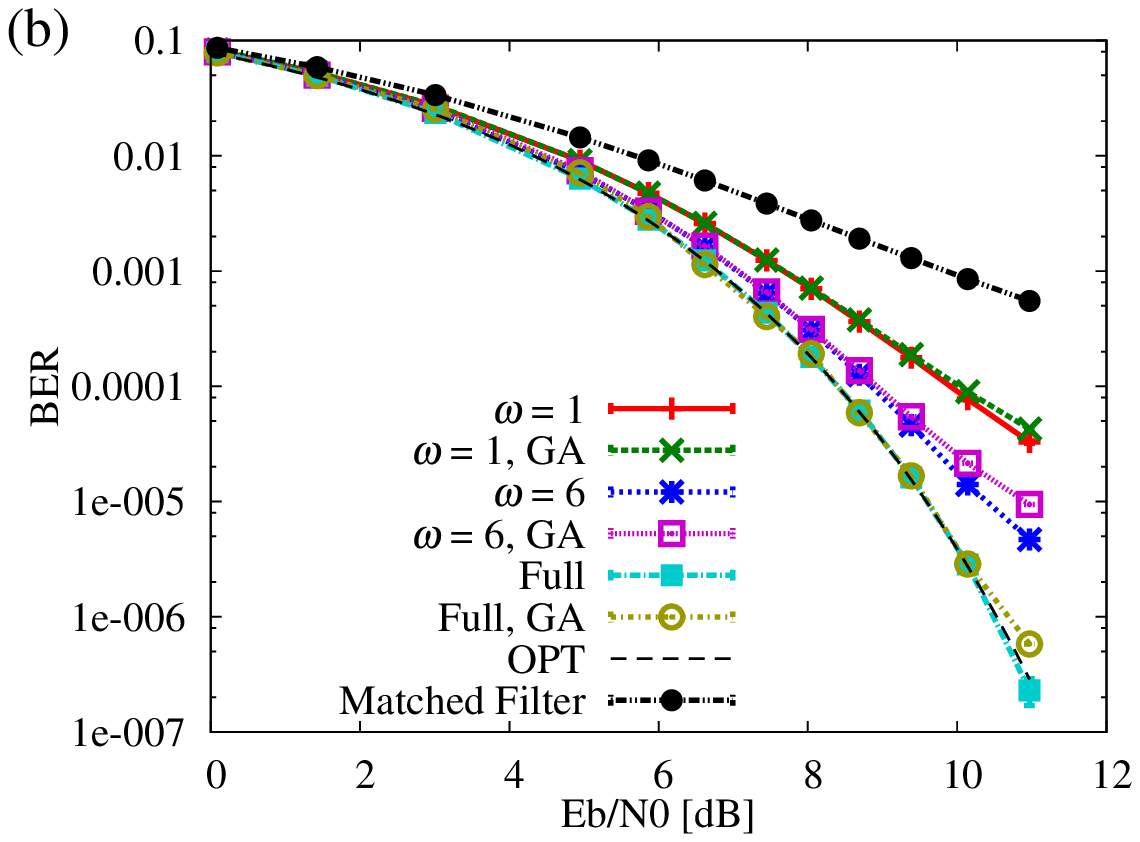}
\end{center}
\caption{(color online) Dependence of BER on Eb$/$N0
at (a) $N=32,~M=3,~\epsilon_{\rm max}=0.5$ and (b) 
 $N=32,~M=15,~\epsilon_{\rm max} = 0.5$.
Results for $\omega=1$ and $\omega=6$ are shown and ``Full'' corresponds to $\omega=16$.
The dashed line and dashed-dotted line represent the 
optimal and matched filter cases, respectively.
GA means the BER performance under the Gaussian approximation.
The data points are averaged over $10^6$ samples of $\bm{W}$.}
\label{figure:BER}
\end{figure}

We check the performance of 
the matched filter, 
whose mathematical manipulation corresponds to multiply the received 
symbol by the Hermit conjugate matrix of $\bm{W}$.
The matched filter minimizes the power of the channel noise, 
but it cannot reduce the error due to the intercarrier interference.
Therefore, as a canceller of intercarrier interference,
the performance of the proposed algorithm 
should be better than that of the matched filter.
In the proposed algorithm, $\omega=0$ corresponds to the demodulation by the matched filter
and as $\omega$ increases, the performance is expected to improve.

Eb$/$N0 dependence of BER is shown in \Fref{figure:BER}
for numbers of the path, (a) $M=3$ and (b) $M=15$, respectively.
The number of the subcarriers is $N=32$, and the maximum value
of the Doppler shift is $\epsilon_{\rm max}=0.5$.
${\rm BER}^{\rm OPT}$ and BER for
the matched filter are shown by the dashed line
and dashed-dotted line, respectively.
The BER performance of the canceller given by
\Eref{decode_w} with $\omega=1$,
which is the simplest case, is better than that of the matched filter,
and it improves as the number of interactions increases.
When all interactions between the subcarriers are 
taken into account,
the BER performance of the canceller given by \Eref{decode_full} 
almost coincides with 
the optimal performance ${\rm BER}^{\rm OPT}$. 

\begin{figure}
\begin{center}
\includegraphics[width=\figwidth]{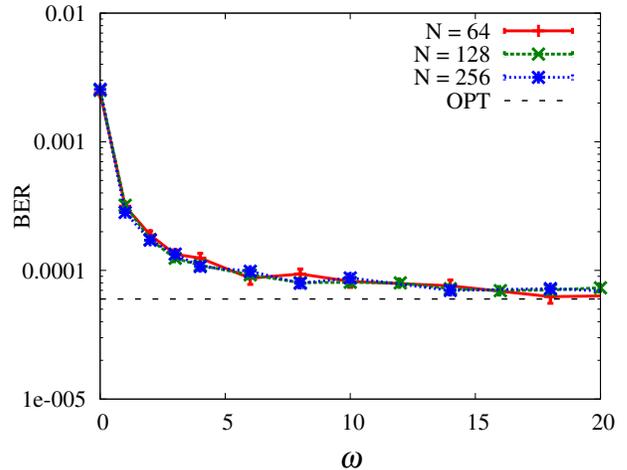}
\end{center}
\caption{(color online) Dependence of the BER performance on $\omega$ 
at $M=3,~\epsilon_{\rm max}=0.5$, and Eb$/$N0$=8.69$dB.
Results for numbers of subcarriers
$N=64,~128$, and 256 are shown,
and the dashed line shows the BER of the 
optimal
limit.}
\label{figure:BER_vs_w}
\end{figure}

The $\omega$-dependence of BER performance at $M=3,~\epsilon_{\rm max}=0.5$, 
and Eb$/$N0$=8.69$ 
is shown in \Fref{figure:BER_vs_w} for $N=64,~128$, and 256. 
This graph indicates that BER does not depend on the number of subcarriers, $N$.
BER rapidly decreases as $\omega$ increases from zero,
and it gradually 
approaches 
${\rm BER}^{\rm OPT}$
as $\omega$ further increases.
At $\omega\simgeq 16$, the differences between the BER of the proposed method
and 
the optimal
limit is less than $10\%$ of the value of 
${\rm BER}^{\rm OPT}$.
This result indicates 
that $\omega\sim 16$ is 
sufficient to practically 
achieve the BER of the original 
MFA based canceller
(\ref{decode_full})
irrespective of $N$; thus, the required numerical 
cost per bit does not increase as $O(N)$. In this model, the unit width of the frequency 
domain is given by $1\slash N$, then the proposed algorithm will provide more effective use of 
the frequency domain without increasing the numerical cost.

\subsection{ICI dynamics}

To determine the validity of the proposed method as a realistic canceller, 
the time evolution of the BER of the proposed algorithm,
decoding algorithm at $N=128,~M=15$, and $\epsilon_{\rm max}=0.5$ 
is plotted in \Fref{figure:BER-dynamics}.
The initial condition, $\{\hat{X}_k^0\}$, is obtained by the matched filter,
so BER at the 0-th step corresponds to that at the fixed point for $\omega=0$.
The horizontal axis means the number of time steps, 
and the vertical axis represents the BER with respect to the tetative
decision at the time step.
As can be seen in the figure, the BER performances at $\omega=3$ and $\omega=64$ (full) converge 
to equilibrium values after two updates.
The time steps required to reach the fixed point are quite short and 
only slightly depend on $N$, 
and the proposed algorithm is useful for an implementation 
as an ICI canceller.

\subsection{Macroscopic description}

To analyze the time evolution of the canceller given by
\Eref{decode_w},
we attempt 
to describe the dynamics by using a finite number of macroscopic 
variables \cite{Tanaka-Okada,Okada}. The overlap between the transmitted bit 
and the predicted bit at step
$t$ under a given realization of $\bm{W}$ is defined as 
\begin{align}
m^t=\frac{1}{2N}\sum_{k=1}^{2N}\langle\hat{X}_k^tX_k\rangle, 
\label{mt_def}
\end{align}
and the 
BER at step $t$ is given by $0.5\times(1-m^t)$. The simplest description 
of the time evolution of \Eref{mt_def}
is provided by ignoring all spatial/time correlations among the subcarrier 
symbols, which leads to a discrete map of $m^t$:
\begin{align}
m^{t+1}&=\int Dz~ {\rm sgn}(z_0+\sqrt{\Sigma_t^2}z),\label{decode_gauss}\\
\nonumber
z_0&=\frac{1}{2N}\sum_{k=1}^{2N}J_{kk},~~~{\Sigma_t^2}=2A(1-m^t)+B+\sigma^2C,
\end{align}
where 
$Dz=dz e^{-z^2\slash 2}/\sqrt{2 \pi}$,
and the coefficients for the fixed sample of $W$ are given by
\begin{align}
A&=\frac{1}{2N}\sum_{k=1}^{2N}\sum_{l\in\partial_k(w)}J_{kl}^2,\\
B&=\frac{1}{2N}\sum_{k=1}^{2N}\sum_{l\nin\partial_k(w)}J_{kl}^2,\\
C&=\frac{1}{2N}\sum_{k=1}^{2N}\sum_{l}W_{lk}^2=z_0.
\end{align}
The derivation is given in detail in Appendix \ref{variance_derivation}.
These expression indicate that the ICI is approximated by Gaussian noise,
and noise variance $\sigma^2$ is effectively increased by $B/C$
due to the insufficiency of the ICI cancellation of
the MFA canceller (\ref{decode_w}). 
The macroscopic equation corresponding to \Eref{decode_full} is obtained by setting $\omega=N\slash 2$.

\begin{figure}
\begin{center}
\includegraphics[width=\figwidth]{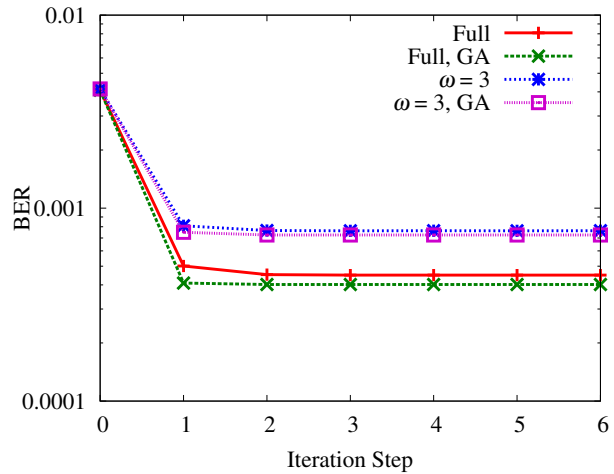}
\end{center}
\caption{(color online) The time-evolution of BER
at $N=128,~M=15,~\epsilon_{\rm max}=0.5$, and Eb$/$N0=7.45dB.
Results for $\omega=3$ and full case $(\omega=64)$ are shown, and
GA means the corresponding Gaussian approximation.
The trajectory is averaged over $10^4$ samples of $\bm{W}$.}
\label{figure:BER-dynamics}
\end{figure}

BER defined at fixed points given by \Eref{decode_gauss},
denoted by GA (Gaussian approximation), are compared to the real BER curve
in \Fref{figure:BER},
where BER $=(1-m_t)\slash 2$.
BER at the fixed point
are in good accordance with the experimental data irrespective of the value of $\omega$. 
The time evolution of BER is also well described by \Eref{decode_gauss}, 
as shown in \Fref{figure:BER-dynamics},
in which the initial condition, $\{m_i^0\}$, is chosen to correspond to 
the BER of the matched filter. 
These results differ substantially from that of the 
random spreading codes, where time correlation plays a significant role 
in the macroscopic dynamics \cite{Tanaka-Okada}. 
The difference implies
that the orthogonality between the subcarriers in OFDM 
reduces the time correlation and enables the 
ICI dynamics to converge within a few steps.

\section{Conclusion}
\label{summary}

A practically feasible ICI canceller for the OFDM model,
which can be regarded as a variant of the
random field spin model,
was developed.
The cancellation scheme was derived by applying 
the mean-field approximation to the maximization 
of the posterior probability which corresponds to 
the search of the ground state of the spin model. 
The properties of the frequency-domain matrix, i.e., translation symmetry 
and smallness of off-diagonal elements compared to diagonal elements, 
was focused on,
and only a part of the ICI among the subcarriers located within a distance $\omega$ 
in the frequency domain for each bit was considered. 
The numerical cost of the ICI canceller
is thus controlled by the parameter 
$\omega$. When $\omega=\lfloor N\slash 2\rfloor$,
the ICI between all subcarriers is considered, and 
the ICI canceller corresponds to the approximated MAP demodulater.

The BER of the proposed algorithm used for QPSK modulation 
is better than the matched filter even if $\omega=1$,
and it practically approaches the 
optimal
limit as $\omega$ increases further. 
The BER performance is saturated near the 
optimal
limit at around a certain $\omega\sim O(1)$ 
in the whole Eb$/$N0 region. Furthermore, the performance under a given $\omega$ 
only slightly depends on $N$.  
This result means that the required numerical cost per iteration to achieve a feasible performance level is $O(N)$. 

The number of time steps required to reach 
a fixed point of ICI canceller is $O(1)$
only slightly depends on the value of $N$ and $\omega$. 
The total numerical cost
to eliminate ICI and demodulate the transmitted bits
is therefore $O(N)$.
The proposed algorithm will be practical to implement 
by virtue of its low computational cost.

The fixed point of the ICI canceller and the dynamics to reach there 
are well described by 
a discrete map of a single macroscopic variable
under the approximation of ICI for 
each bit as independent Gaussian noise.
It is considered that the orthogonality
between the subcarriers prevents a
time correlation being induced,
and is a mathematical background of the 
high accuracy of the 
proposed ICI canceller.

The proposed algorithm is efficient 
for the parameter region where 
the Doppler shift causes large intercarrier interference.
By introducing this algorithm as an ICI canceller,
the OFDM scheme is useful when subcarriers are closely arraigned in the frequency 
domain and a mobile object moves at high speed. 
More efficient use of a given frequency 
domain and enhanced accuracy in satellite communications are also expected.


\begin{acknowledgments}
We would like to thank Ido Kanter for his helpful comments and discussions. 
This work was supported by JSPS Fellow No. 23--4665 (AS)
and KAKENHI No. 22300003 (YK).
\end{acknowledgments}

\appendix

\section{Derivation of macroscopic dynamics}
\label{variance_derivation}

According to the definition of $\{h_k\}$, \Eref{decode_w} can be transformed as follows,
\begin{align}
\hat{X}_i^{t+1}={\rm sgn}[-\sum_{j\in\partial_i(\omega)}J_{ij}\hat{X}_j^t+\sum_{j=1}^{2N}J_{ij}X_j+\sum_{\mu = 1}^{2N}W_{\mu i}n_\mu].
\end{align}
By separating the summation of the first and second term into three part;
$j\in\partial_i(\omega)$, $j\ninn\partial_i(\omega)$ and $j=i$,
the time evolution of $m_i^t$ can be written as 
\begin{align}
\nonumber
m_i^{t+1}&=\langle{\rm  sgn}[J_{ii}+X_i\sum_{j\in\partial_i(\omega)}J_{ij}(X_j-\hat{X}_j^t)\\
\nonumber
&~~~~~+X_i\sum_{j\nin\partial_i(\omega)}J_{ij}X_j+X_i\sum_{\mu}W_{\mu i}n_\mu]\rangle\\
&\equiv\langle{\rm sgn}[J_{ii}+X_iz_i^t]\rangle.
\label{appendix:m_t-1}
\end{align}
The right-hand side of \Eref{appendix:m_t-1} depends on the randomness;
the transmitted symbol and the channel noise, through $z_i^t$.
The average over the randomness can therefore be replaced by the
average over $z_i^t$ according to an appropriate distribution. 
The distribution of $z_i^t$ is approximated by a Gauss distribution.
The first and second moments of $z_i^t$ are given by
\begin{align}
\langle z_i^t\rangle&=0,\\
\langle {z_i^t}^2\rangle&=2\sum_{j\in\partial_i(\omega)}J_{ij}^2(1-m_j^t)+2\sum_{j\nin\partial_i(\omega)}J_{ij}^2+\sigma^2\sum_{\mu=1}^{2N}W_{\mu i}^2,
\end{align}
where it is assumed that $X_j=\hat{X}_j^t$ with probability 
$(1+m_j^t)\slash 2$, and $X_j=-\hat{X}_j^t$ with probability
$(1-m_j^t)\slash 2$ for any $j$.
With these quantities, the following expression can be obtained:
\begin{align}
m_i^t=\int Dz~{\rm sgn}[J_{ii}+X_i\sqrt{{\Sigma_i^t}^2}z],
\label{appendix:m_t-2}
\end{align}
where ${\Sigma_i^t}^2=\langle {z_i^t}^2\rangle-\langle {z_i^t}\rangle^2$, and
coefficient $X_i$ can be ignored because the function is 
invariant against translation $z\to -z$.
The approximated expression for the full case can be obtained
by setting $\omega=N\slash 2$, so $\partial_i(\omega)$ contains all bits except $i$.

\newpage 

\end{document}